\documentclass[twocolumn,twoside]{IEEEtran}

\usepackage{mathpple}
\usepackage{times}

\usepackage{amsmath}
\usepackage{amsfonts}
\usepackage{latexsym}
\usepackage{amssymb}

\usepackage{upref}
\usepackage{theorem}

\usepackage{graphicx}
\usepackage{psfrag}

\title{$\,$\\[-1.5ex]
On the Stopping Distance and the Stopping~Redundancy of Codes\\[0.05ex]}

\author{\large 
Moshe~Schwartz,~\IEEEmembership{\large Member,~IEEE,}
and Alexander Vardy,~\IEEEmembership{\large Fellow,~IEEE}
\vspace{-4ex}
\thanks{Manuscript submitted March 16, 2005; revised December 9, 2005.
   This work was supported in 
   part~by the National Science Foundation 
   and in part by the David and Lucile Packard~Foundation.
   The material in this paper was presented in part 
   at the IEEE International Symposium on Information Theory, 
   Adelaide, Australia, September 2005.
}
\thanks{Moshe Schwartz was with the 
   Department of Electrical and Computer~Engineering,
   University of California San Diego,
   and is now with the Depart\-ment~of Electrical Engineering,
   California Institute of Technology,
   1200 East California Blvd., Mail Code 136-93,
   Pasadena, CA 91125, U.S.A.
   (e-mail: moosh@paradise.caltech.edu).}
\thanks{Alexander Vardy is with the 
   Department of Electrical and Computer~Engi\-neering,
   the Department of Computer Science and Engineering,
   and the Depar\-tment of Mathematics,
   all at the University of California San Diego,
   La Jolla, CA 92093--0407, U.S.A.
  (e-mail: vardy@kilimanjaro.ucsd.edu).}
}

\renewcommand{\markboth}[2]
{\renewcommand{\leftmark}{#1}\renewcommand{\rightmark}{#2}}

\theoremstyle{plain} 
\theorembodyfont{\normalfont\slshape}

\newtheorem{thm}{Theorem$\!$} 
\newenvironment{theorem}
{\begin{thm}\hspace*{-1ex}{\bf.}}{\end{thm}}

\newtheorem{lem}[thm]{Lemma$\!$}
\newenvironment{lemma}{\begin{lem}\hspace*{-1ex}{\bf.}}{\end{lem}}

\newtheorem{prop}[thm]{Proposition$\!$}
\newenvironment{proposition}{\begin{prop}\hspace*{-1ex}{\bf.}}{\end{prop}}

\newtheorem{cor}[thm]{Corollary$\!$}
\newenvironment{corollary}{\begin{cor}\hspace*{-1ex}{\bf.}}{\end{cor}}

\newtheorem{defn}{Definition$\!$}
\newenvironment{definition}{\begin{defn}\hspace*{-1ex}{\bf.}}{\end{defn}}

\newcounter{enumrom}
\renewcommand{\theenumrom}{(\roman{enumrom})}

\makeatletter
\renewcommand{\@endtheorem}{\endtrivlist}
\makeatother

\makeatletter
\renewcommand{\thefigure}{{\bf \@arabic\c@figure}}
\renewcommand{\fnum@figure}{{\bf Figure}\,\thefigure}
\makeatother

\renewcommand{\QEDclosed}{\mbox{\rule[-1pt]{1.3ex}{1.3ex}}} %

\newcommand{\cC}{{\cal C}} 
\newcommand{\cD}{{\cal D}}
 
\newcommand{\cF}{{\cal F}}
\newcommand{\cG}{{\cal G}} 
\newcommand{\cH}{{\cal H}}
\newcommand{\cI}{{\cal I}}

\newcommand{\cS}{{\cal S}}

\newcommand{\mathset}[1]{\left\{#1\right\}}
\newcommand{\abs}[1]{\left|#1\right|}
\newcommand{\ceilenv}[1]{\left\lceil #1 \right\rceil}
\newcommand{\floorenv}[1]{\left\lfloor #1 \right\rfloor}
\newcommand{\parenv}[1]{\left( #1 \right)}

\newcommand{\bracenv}[1]{\left\{ #1 \right\}}

\newcommand{\be}[1]{\begin{equation}\label{#1}}
\newcommand{\ee}{\end{equation}} 
\newcommand{\eq}[1]{(\ref{#1})}

\renewcommand{\le}{\leqslant} 
\renewcommand{\leq}{\leqslant}
\renewcommand{\ge}{\geqslant} 
\renewcommand{\geq}{\geqslant}

\renewcommand{\Bbb}{\mathbb}

\newcommand{\bfsl}{\bfseries\slshape}

\newcommand{\Tref}[1]{Theo\-rem\,\ref{#1}}
\newcommand{\Pref}[1]{Pro\-po\-si\-tion\,\ref{#1}}
\newcommand{\Lref}[1]{Lem\-ma\,\ref{#1}}
\newcommand{\Cref}[1]{Co\-ro\-lla\-ry\,\ref{#1}}

\renewcommand{\Bbb}{\mathbb}

\newcommand{\Fq}{{{\Bbb F}}_{\!q}}
\newcommand{\Ftwo}{{{\Bbb F}}_{\!2}} 
\newcommand{\Fn}{\Bbb{F}_{\!2}^{\hspace{1pt}n}}
\newcommand{\Ffour}{{{\Bbb F}}_{\!4}}

\newcommand{\C}{{\Bbb C}}

\newcommand{\deff}{\mbox{$\stackrel{\rm def}{=}$}}

\newcommand{\Strut}[2]{\rule[-#2]{0cm}{#1}}

\newcommand{\sd}{s}
\newcommand{\grm}{{\cal{R}}}
\newcommand{\Gt}{H}
\newcommand{\gtop}{\Gt_{\text{top}}}
\newcommand{\gbot}{\Gt_{\text{bot}}}

\DeclareMathOperator{\wt}{wt}
\DeclareMathOperator{\rank}{rank}
\DeclareMathOperator{\supp}{supp}

\newcommand{\zero}{{\mathbf 0}}
\newcommand{\one}{{\mathbf 1}}

\newcommand{\oH}{\overline{H}}

\DeclareMathAlphabet{\mathbfsl}{OT1}{cmr}{bx}{it}
\newcommand{\hhh}{\mathbfsl{h}}

\newcommand{\NML}{\smash{\Psi_{\rm ML}(w)}}
\newcommand{\NH}{\smash{\Psi_{H_{24}}\hspace{-0.25ex}(w)}}
\newcommand{\NHprime}{\smash{\Psi_{H'_{24}}\hspace{-0.35ex}(w)}}
\newcommand{\NMLi}{{\Psi_{\rm ML}}}
\newcommand{\NHi}{{\Psi_{H_{24}}}}
\newcommand{\NHprimei}{{\Psi_{H'_{24}}}}
\newcommand{\NHt}{\smash{\Psi_{H_{12}}\hspace{-0.25ex}(w)}}
\newcommand{\NHprimet}{\smash{\Psi_{H'_{12}}\hspace{-0.35ex}(w)}}

\newcommand{\w}{{\omega}}
\newcommand{\bw}{{\overline{\omega}}}

\outer\def\proclaim #1. #2\par{\medbreak
 \noindent{\bf#1.\enspace}{\sl#2\par}%
 \ifdim\lastskip<\medskipamount \removelastskip\penalty55\medskip\fi}

\mathchardef\inn="3232
\renewcommand{\in}{\mbox{$\,\inn\,$}}

\begin{document}

\maketitle

\begin{abstract}
It is now well known that the performance of~a~linear code $\C$
under iterative decoding on a binary erasure channel (and other channels) 
is determined by the size of the smallest stopping set in 
the Tanner graph for $\C$. Several recent papers~refer to 
this parameter as the \emph{stopping distance} $s$ of $\C$. 
This~is~some-what of a misnomer since the size of the smallest 
stopping set in the Tanner graph for $\C$ depends on the 
corresponding choice~of a parity-check matrix. It is easy 
to see that $s \le d$,~where~$d$~is~the minimum Hamming 
distance of $\C$, and we show that it is always possible
to choose a parity-check matrix for $\C$ (with sufficiently
many dependent rows) such that $s = d$. We thus introduce~a~new
parameter, termed the \emph{stopping redundancy} of $\C$,
defined~as~the minimum number of rows in a parity-check
matrix $H$ for $\C$ such that the corresponding stopping 
distance $s(H)$ attains its largest possible value, namely $s(H) = d$.
We then derive general bounds on the stopping
redundancy of linear codes. We also examine several simple 
ways of constructing codes from other codes,~and
study the effect of these constructions on the stopping~redundan-cy. 
Specifically, for the family of binary Reed-Muller codes (of all orders),
we prove that their stopping redundancy is at most a~constant times 
their conventional redundancy. We show that the stopping redundancies
of the binary and ternary extended Golay codes are at most $34$ and $22$, 
respectively. Finally, we provide~up- per and lower bounds on the
stopping redundancy of MDS codes\mbox{.\hspace*{-1pt}}\vspace{-2ex} 
\end{abstract}

\begin{keywords}
erasure channels,
Golay codes,
iterative~decoding,
linear codes, 
MDS codes,
Reed-Muller codes,
stopping sets.\vspace{-1ex}
\end{keywords}

\section{Introduction} 
\vspace{-.25ex}
\label{sec:Introduction}

\noindent\looseness=-1
\PARstart{T}{he} recent surge of interest in the binary erasure
channel (BEC) is due in large part to the fact that~it~is~the~prime 
example of a channel over which the performance of 
iterative decoding algorithms can be analyzed precisely. 
In particular, it was shown by 
Di, Proietti, Telatar, Richardson,~and~Urb\-anke~\cite{DPTRU}
that the performance of an LDPC code (and,~in~fact, any linear
code) under iterative decoding on the~BEC~is~completely 
determined by certain combinatorial structures
called \emph{stopping sets}. A stopping set $\cS$ 
in a code $\C$ is a~subset~of~the variable nodes in a
Tanner graph for $\C$ such that~all~the~neigh-bors of $\cS$ 
are connected to $\cS$ at least twice. The size $s$ of the 
smallest stopping set was termed the \emph{stopping distance}
of~$\C$ in a number of recent papers~\cite{KV03,OVZ}.
The stopping distance plays an important role in understanding
the performance of a~code under iterative decoding over the BEC,
akin to the role played by the minimum Hamming distance~$d$ 
for maximum-likelihood and/or algebraic decoding. Just as one would 
like to maximize the minimum distance $d$ if maximum-likelihood 
or algebraic decoding is to be used, so one~should~try to maximize 
the stopping distance $s$ in the case of iterative decoding.

There is, however, an important difference between~the~min\-i\-mum 
distance $d$ and the stopping distance $s$. While the former is
a~property of a code $\C$, the latter depends on the specific 
Tanner graph for $\C$ or, equivalently, on the specific 
choice of a parity-check matrix $H$ for $\C$. In order to 
emphasize this, we will henceforth use 
$s(H)$ to denote the stopping distance
and 
$d(\C)$ to denote the minimum distance.

\looseness=-1
In algebraic coding theory, a parity-check matrix $H$ 
for~a~linear code $\C$ usually has $n - \dim(\C)$ linearly
independent~rows. However, in the context of iterative decoding, it has 
been~alre\-ady observed in~\cite{SV04},\,\cite{YCF}
and other papers that adding linearly dependent rows to $H$
can be advantageous. Certainly, this can increase the stopping
distance $s(H)$. Thus, throughout this~pa\-per, a 
\emph{parity-check matrix} for $\C$ should be understood~as~any 
matrix $H$ whose rows span the dual code $\C^\perp$.
Then the~\emph{redundancy} $r(\C)$ of $\C$ may be defined
as the minimum number of rows in a parity-check matrix for $\C$.
Analogously, we define the \emph{stopping redundancy} 
$\rho(\hspace{-.5pt}\C)$ of $\C$ as the minimum number 
of rows in a parity-check 
matrix $H$ for $\C$ such that $s(H) = d(\C)$. This work may be
thought of as the first investigation of the trade-off between
the parameters $\rho(\C)$, $r(\C)$, and $d(\C)$.

In the next section, we first show that the stopping 
redundancy $\rho(\C)$ is well-defined. That is, given
any linear code~$\C$, it is always possible to find 
a parity-check matrix $H$ for $\C$ such that $s(H) = d(\C)$. 
In fact, the parity-check matrix consisting of \emph{all}
the nonzero codewords of the dual code $\C^\perp$ has
this property. Hence $\rho(\C) \le 2^{r(\C)} - 1$ for
all binary linear codes. 
We then show in Section\,II
that if $d(\C) \le 3$, then \emph{any} parity-check 
matrix $H$ for $\C$ satisfies $s(H) = d(\C)$, so
$\rho(\C) = r(\C)$ in this case. 
The main result of Section\,II is an extension of this 
simple observation to a \emph{general upper bound} on the 
stopping redundancy of binary linear codes (\Tref{general}). 
We also derive in Section\,II~a general \emph{lower bound}
on the stopping~redundancy 
of linear codes (\Tref{lowerbound}). 

In Section\,III, we study several simple ways of constructing
codes from other codes, such as the direct-sum construction
and code extension by adding an overall parity-check. We 
investigate the effect of these constructions on the stopping
redundancy.
It should be pointed out that although we have
focused our discussion on binary codes in Sections II and III,
most of the results therein extend straightforwardly
to linear codes over an arbitrary finite field.

We continue in Section\,IV with an in-depth 
analysis~of~the well-known $(u,u+v)$ construction, 
and in particular~its~application in the recursive
definition~\cite[p.\,374]{MWS} of binary Reed-Muller codes. By slightly
modifying this construction,~we~establish a~strong
upper bound on the stopping redundancy of Reed-Muller codes of
arbitrary orders. 
Specifically, we prove that if $\C$ is a Reed-Muller
code of length $2^m$ and order $r$, then
$
\rho(\C) \le d(\C) r(\C) / 2
$.
Thus for any constant $d(\C)$, we have an increase 
in redundancy by only a constant factor.

\looseness=-1
In Section\,V, we study the $(24,12,8)$ extended binary~Golay code
$\cG_{24}$ and the $(12,6,6)$ extended ternary Golay code $\cG_{12}$. 
We prove that 
$\rho(\cG_{24}) \le 34$ and $\rho(\cG_{12}) \le 22$ by 
providing specific parity-check matrices for these codes. We take 
$\cG_{24}$~as a test case, and compare the performance of three 
different decoders: a maximum-likelihood decoder, an iterative decoder 
using the conventional $12 \times 24$ double-circulant parity-check 
matrix of \cite[p.65]{MWS}, and an iterative decoder using the 
$34\,{\times}\,24$ 
parity-check matrix with maximum stopping distance.
In each case, exact analytic expressions for the probability 
of decoding failure are derived using a computer program (see Figure\,1). 

In Section\,VI, we consider MDS codes over $\Fq$ with $q \ge 2$.
It is easy to extend the general bounds of Section\,II to $q$-ary codes.
However, in Section\,VI we establish much better upper and lower bounds
on the stopping redundancy of MDS codes. 
Notably, all these bounds are independent of the field size $q$. 

This paper only scratches the surface of the many interesting and
important questions that arise in the investigation of the stopping 
redundancy. We conclude in Section\,VII with a brief discussion 
and a~list of open problems.

\vspace{2ex}
\section{General Bounds}
\vspace{-.25ex}
\label{sec:general}

\noindent
We begin with rigorous definitions of the stopping distance and 
the stopping redundancy. 
Let $\C$ be a binary linear code and let $H = [h_{i,j}]$ be 
a parity-check matrix for $\C$. The corresponding Tanner graph
$T$ for $\C$ is a bipartite graph with each column of $H$ 
represented by a \emph{variable node} and each row of 
$H$~re\-presented by a \emph{check node} in such a way that the
$j$-th variable node is connected to the $i$-th check node
if and only if \mbox{$h_{i,j} \ne 0$}.
As already mentioned, a stopping set in $T$ is 
a~subset~$\cS$~of  the variable nodes such that all
the check nodes that are neigh\-bors of a node in $\cS$ are connected 
to \emph{at least two nodes} in $\cS$. We dispense with this graphical
representation of stopping sets~in favor of an equivalent definition 
directly in terms of~the~under\-lying parity-check matrix $H$. 
Thus we say that a \emph{stopping set} is a set of columns 
of $H$ with the property that the projection of $H$ onto 
these columns does not contain a row of weight~one\footnote{%
%
%
\looseness=-1
This explains why stopping sets \emph{stop} the progress of 
an iterative decoder. 
A row of weight one --- equivalently,
a check node of degree one --- would determine unambiguously an
erased symbol. However, if an entire stopping
set is erased, then all the neighboring check nodes are connected to 
these erasures at least twice, and thus form an under-constrained system
of linear equations. 
In this case, an iterative decoder has no way 
of determining the erased values.
}.
The resulting definition of the stopping distance -- the smal\-lest size 
of a stopping set -- bears a striking resemblance~to~the~de\-fini\-tion 
of the minimum Hamming distance of a linear code.

Recall that the minimum distance of a linear code $\C$~can~be 
defined as the largest integer $\smash{d(\C)}$ 
such that 
every $\smash{d(\C)-1}$ or less columns of $H$ are linearly 
independent. 
For binary codes, this is equivalent to saying that
$d(\C)$ is the largest integer 
such that 
every set of $d(\C)-1$ or less columns of $H$ {contains 
at least one \underline{row of odd weight}}.

\begin{definition}
Let\/ $\C$ be a linear code (not necessarily binary)
and let $H$ be a parity-check matrix 
for\/ $\C$. Then the {\bfsl stopping distance} of $H$ is defined
as the the largest integer $s(H)$ 
such that 
every set of\/ $s(H)-1$ or less columns of $H$ {contains 
at least one \underline{row of weight one}}.
\end{definition}

\hspace*{-.8pt}The following corollary is an immediate 
consequence~of~jux\-taposing the definitions of $s(H)$ and $d(\C)$ above.

\begin{corollary}
Let\/ $\C$ be a linear code and let $H$ be an arbitrary
parity-check matrix for\/ $\C$. Then $s(H) \leq d(\C)$.
\end{corollary}

\noindent
Indeed, it is well known~\cite{DPTRU,Feldman,KV03} that the 
support of every codeword is a stopping set, which is another
way to see that $s(H) \leq d(\C)$ regardless of the choice of $H$.
Thus given~a~lin\-ear code $\C$, the largest stopping distance one
could hope for is~$d(\C)$, no matter how cleverly the Tanner
graph for $\C$~is~con\-structed. The point is that this bound
can be \emph{always} achieved 
by adding dependent rows to $H$
(see \Tref{trivial}).
This makes the notion of the stopping distance, as 
a property of a code $\C$, somewhat meaningless:
without restricting the number of rows in a parity-check
matrix for $\C$, we cannot distinguish between the 
stopping distance and the conventional minimum distance.
This observation, in turn, leads to the following
definition.

\begin{definition}
Let\/ $\C$ be a linear code 
with minimum Hamming distance $d(\C)$. 
Then the {\bfsl stopping redundancy} 
of\/ $\C$ is defined as the the smallest integer $\rho(\C)$ 
such that there exists a parity-check matrix $H$ for\/ $\C$
with $\rho(\C)$ rows and $s(H) = d(\C)$.
\end{definition}

The following theorem shows that the stopping redundancy 
is, indeed, well-defined.

\begin{theorem}
\label{trivial}
Let\/ $\C$ be a linear code, and let $H^*$ denote 
the~pari-ty-check matrix for\/ $\C$
consisting of all the nonzero codewords of the dual 
code\/ $\C^\perp$. Then $s(H^*) = d(\C)$.
\vspace{-.3ex}
\end{theorem}

\begin{proof}
Let \looseness=-1
$[\C^\perp]$ 
denote the $|\C^\perp| \times n$ matrix
consisting of all the codewords of $\C^\perp$. It is 
well known (cf.~\cite[p.139]{MWS}) that $[\C^\perp]$ 
is an orthogonal array of strength $d(\C)-1$. This 
means that any set of $t \le d(\C)-1$ columns of
$[\C^\perp]$ 
contains all the vectors of length $t$ among its 
rows, each vector appearing the same number of times.
In particular, any set of $d(\C){-}1$~or~less columns of 
$[\C^\perp]$ 
contains all the vectors of weight one among its rows.
Clearly, the all-zero row can be removed from 
$[\C^\perp]$ to obtain $H^*\!$, while
preserving this property.
\vspace{1ex}
\end{proof}

\looseness=-1
\Tref{trivial} also provides a trivial 
upper~bound~on~the~stop\-ping redundancy. 
In particular, it follows 
from~\Tref{trivial}~that 
$\rho(\C) \le 2^{r(\C)} - 1$ for any binary linear
code $\C$. This bound holds with equality in 
the degenerate case of the single-parity-
check code. The next theorem determines
$\rho(\C)$ exactly for~\emph{all} binary
linear codes with minimum distance $d(\C)\! \le 3$.

\begin{theorem}
\label{dist3}
Let\/ $\C$ be a binary linear code with minimum~dist\-ance 
$d(\C)\! \le 3$. Then {\bfsl any} parity-check matrix $H$ for\/ $\C$
satisfies $s(H) = d(\C)$, and therefore $\rho(\C)=r(\C)$.
\vspace{-.3ex}
\end{theorem}

\begin{proof}
If $H$ contains an all-zero column, then it is obvious that
$s(H) = d(\C) = 1$.
Otherwise $s(H) \ge 2$, since then every single column of $H$ must contain
a row of weight one. Now, if $d(\C) = 3$, then every two columns of $H$
are distinct. This implies that these two columns must contain either
the $01$ row or the $10$ row (or both). Hence $s(H) = 3$.
\vspace{1ex}
\end{proof}

\looseness=-1
The following theorem, which is our main result in this section,
shows that \Tref{dist3} is, in fact, a special case of a general
upper bound on the stopping redundancy of linear codes.

\begin{theorem}
\label{general}
Let\/ $\C$ be a binary linear code with minimum~dist\-ance 
$d(\C)\! \ge 3$. Then
\be{thm4}
\rho(\C)
\, \le \,
\binom{r(\C)}{1} + \binom{r(\C)}{2} + \cdots + \binom{r(\C)}{d(\C)-2} 
\ee
\end{theorem}

\begin{proof}
We first prove a slightly weaker result,~which~is~con\-ceptually
simpler. Namely, let us show that 
\be{thm4-aux}
\rho(\C)
\, \le \,
\binom{r(\C)}{1} + \binom{r(\C)}{2} + \cdots + \binom{r(\C)}{d(\C)-1} 
\ee
Let $H$ be an arbitrary parity-check matrix for $\C$ with $r(\C)$~linearly 
independent rows. Construct another parity-check matrix $H'$ whose 
rows are all the linear combinations of $t$ rows~of~$H$,
for all $t = 1,2,\ldots,d(\C)-1$. Clearly, the number of 
rows~of $H'$ is given by the right-hand side of \eq{thm4-aux}.
Now let $H_t$,~respectively $H'_t$, denote a matrix consisting
of some $t$ columns of $H$, respectively the corresponding $t$
columns of $H'$. Observe~that 
for all $t \le d(\C)-1$, the $t$
columns of $H_t$ are linearly~indepen-dent. This implies that 
the row-rank of $H_t$ is $t$, and therefore some $t$ rows of 
$H_t$ must form a basis for $\Ftwo^t$. Hence~the~$2^t-1$ 
nonzero linear combinations of these $t$ rows of $H_t$ generate all the
nonzero vectors in $\Ftwo^t$, including all the vectors of weight 
one. But for $t \le d(\C)-1$, the $2^t-1$ nonzero linear~combin-
ations of \emph{any} $t$ rows of $H_t$ are among the rows of $H'_t$
by~con-struction. This proves that $s(H') = d(\C)$ and 
establishes~\eq{thm4-aux}.~~

To transition from \eq{thm4-aux} to \eq{thm4}, observe that we
do not need~to have all the nonzero vectors of $\Ftwo^t$ among
the rows of $H'_t$; it would suffice to have at least one
vector of weight one. Given a set $\cS \subseteq \Ftwo^t$ and
a positive integer $m$, let $m\cS$ denote the set of all vectors 
obtained as a linear combination of at most $m$ vectors from $\cS$.
Define $\mu(t)$ as the smallest integer with the property that 
for any basis $B$ of $\Ftwo^t$, the set $\mu(t)B$ contains at 
least one vector of weight one. Then in the construction~of~$H'$,
it would suffice to take all the linear combinations of at most
$\mu(d(\C)-1)$ rows of $H$. Clearly $\mu(t) \le t-1$ for all $t$
(in~fact, $\mu(t) = t-1$ for all $t$), and the theorem follows.
\vspace{1ex}
\end{proof}

The bound of \eq{thm4}, while much better than 
$\rho(\C) \le 2^{r(\C)} - 1$, is still too general to
be tight for most codes. Nevertheless, we can conclude
from \Tref{general} that when $d(\C)$ is a constant,
the stopping redundancy is only polynomial in the 
(conventional) redundancy and, hence, in the length 
of the code.
\vspace{1ex}

In the next theorem, we provide a general 
\emph{lower bound} on the stopping redundancy 
of linear codes.

\begin{theorem}
\label{lowerbound}
Let\/ $\C$ be an arbitrary linear code of length $n$. 
For each\/ $i = 1,2,\ldots,d(\C)\,{-}\,1$, define
\be{w_i}
w_i
\ \ \deff\
\max\bracenv{\ceilenv{\frac{n+1}{i}}-1,\, d(\C^\perp)}
\vspace{-1ex}    
\ee
Then 
$$
\rho(\C)
\, \geq \,
\binom{n}{i} \!\left/  w_i\binom{n{-}w_i}{i-1} \right.
\hspace{4ex}
\text{for\, $i = 1,2,\ldots,d(\C)\,{-}\,1$}
\vspace{2ex}    
$$
\end{theorem}

\begin{proof}
Let $H$ be a parity-check matrix for $\C$ and let $\cI$ be 
an arbitrary set of $i$ column indices. We say that $\cI$ 
is an \emph{$i$-set}. We~also say that a row $\hhh$ of $H$
\emph{covers} $\cI$ if the projection~of~$\hhh$ onto $\cI$
has weight one. If $s(H) = d(\C)$, then each of the~$\binom{n}{i}$
$i$-sets must be covered by at least one row of the parity-check 
matrix, for all $i = 1,2,\ldots,d(\C)\,{-}\,1$. Any single 
row of $H$~of weight $w \le n-i+1$ covers exactly
\be{f_n}
f_n(i,w)
\,\ \deff \,\
w \binom{n{-}w}{i-1}
\ee
$i$-sets. It is not difficult to see that the expression
in \eq{f_n}~increases monotonically as $w$ decreases until
$f_n(i,w)$ reaches~its maximum at
$
w = \ceilenv{(n+1)/i}-1
$.  
But $\wt(\hhh) \ge d(\C^\perp)$ for all rows $\hhh$ of $H$.
Thus each row of $H$ covers at most $f_n(i,w_i)$ $i$-sets,
where $w_i$ is defined in \eq{w_i}, and the theorem follows.\vspace{1.25ex}
\end{proof}

Is there an asymptotically good sequence of linear codes 
$\C_1,\C_2,\ldots$ such that the stopping redundancy $\rho(\C_i)$
grows only polynomially fast with the length?
The answer to this question is unknown at the present time.
However, if the 
dual sequence
$\C_1^\perp,\C_2^\perp,\ldots$ is also asymptotically good, 
we can use \Tref{lowerbound} to settle this question in the 
negative.

\begin{corollary}
\label{good-dual}
Let\/ $\C_1,\C_2,\ldots$ be an infinite sequence of linear 
codes of strictly increasing length $n_i$ and fixed rate $k_i/n_i = R$,
with $0 \,{<}\, R \,{<}\, 1$, such that $d(\C_i)/n_i \ge \delta_1$
for all $i$, with~$\delta_1\,{>}\,0$.
If also $d(\C_i^\perp)/n_i \ge \delta_2$ for all $i$,
with $\delta_2 > 0$, then 
$$
\rho(\C_i)
\ = \
\Omega\!\left(%
n^{-1}2^{n\left[H_2(\delta_1) \,-\, 
(1-\delta_2)H_2\left(%
\!\frac{\delta_1\Strut{0ex}{0.75ex}}{1-\delta_2\Strut{1.250ex}{0ex}}\!
\right)\right]} 
\right)
$$
where $n = n_i$ and 
\mbox{$H_2(x) = x \log_2 \!x^{-1} \hspace{-1pt}+ (1{-}x) \log_2(1{-}x)^{-1}$}
is the binary entropy function.
\end{corollary}

\begin{proof}
We apply the bound of \Tref{lowerbound} with the size of an $i$-set
given by $d(\C) \,{-}\, 1$.
It is easy to see that if $d(\C_i)/n_i \ge \delta_1$ and
$d(\C_i^\perp)/n_i \,{\ge}\, \delta_2$~for all $i$, then the 
maximum in \eq{w_i}~is~attai\-ned at $\smash{d(\C_i^\perp)}$ for 
all sufficiently large $i$. Thus
\begin{eqnarray*}
\rho(\C_i)
&\hspace{-1.20ex}\geq\hspace{-1.25ex} &
\binom{n}{d(\C_i){-}\,1}%
\!\left/
d(\C_i^\perp)\binom{n\,{-}\,d(\C_i^\perp)}{d(\C_i)-2}
\right.
\\[1.25ex]
&\hspace{-1.5ex}\geq\hspace{-1.5ex} &
\frac{2^{nf(\delta_1,\delta_2)}}{n}
\cdot
\frac{\delta_1}{\delta_2}
\sqrt{\frac{\pi(1{-}\delta_1{-}\delta_2)}{4(1{-}\delta_1)(1{-}\delta_2)}}
\\[-1.25ex]
\end{eqnarray*}
where 
$
f(\delta_1,\delta_2)
=
H_2(\delta_1) - (1{-}\delta_2)
H_2\bigl(\delta_1/(1{-}\delta_2)\bigr)
$
and~the second inequality follows from well-known 
bounds~\cite[p.309]{MWS} on binomial coefficients
in terms of $H_2(\cdot)$. 
\vspace{1.00ex}
\end{proof}

We observe that the function $f(\delta_1,\delta_2)$
defined in the proof of \Cref{good-dual} is always positive, 
and therefore~$\rho(\C_i)$ indeed grows exponentially with the 
length $n$. Note that several well-known families of asymptotically 
good codes (for example,~the self-dual codes~\cite{MST}) 
satisfy the condition of \Cref{good-dual}.

\vspace{2.0ex}
\section{Constructions of Codes from Other Codes}
\label{sec:constructions}

\noindent
In this section, we examine several simple ways of constructing 
codes from other codes. While for most such constructions, it 
is trivial to determine the redundancy of the resulting code, 
we find it considerably more difficult to determine the resulting
{stopping redundancy}, and resort to bounding it. 

\looseness=-1
\hspace{-1pt}We start with two simple examples. 
The first~example~(\Tref{u,v})
is the well-known direct-sum construction
or, equivalently, the $(u,v)$ construction.
The second one (\Tref{u,u}) is the $(u,u)$
construction, or 
concatenation of a code with itself. 

\begin{theorem} 
\label{u,v}
Let\/ $\C_1,\C_2$ be $(n_1,k_1,d_1),(n_2,k_2,d_2)$~binary~linear 
codes, respectively. Then\,
$
\C_3 = \{ (u,v) : u\in\C_1, v\in\C_2 \}
$
is an $(n_1+n_2,k_1+k_2,\min\{d_1,d_2\})$ code with
\be{thm5}
\rho(\C_3)
\,\leq\,
\rho(\C_1) + \rho(\C_2)
\ee
\end{theorem}

\begin{proof}
Let $H_1$ be an arbitrary $\rho(\C_1)\!\times n$ parity-check~matrix 
for $\C_1$ with $s(H_1) = d_1$, and
let $H_2$ be an arbitrary $\rho(\C_2)\times n$ parity-check matrix for $\C_2$
with $s(H_2) = d_2$. 
Then
$$
H_3
\,=\,
\left(
\begin{array}{@{}c@{\ }c@{}}
H_1   & \zero \\
\zero & H_2
\end{array}
\right)
$$
is a parity-check matrix for $\C_3$. Assume w.l.o.g.\ that
$d_1\leq d_2$, so $d(\C_3) = d_1$. Label the
columns of $H_3$ by $1,2,\ldots,n_1+n_2$,
and let $\cI$ be an arbitrary set of at most $d(\C_3)-1$ column~ind\-ices.
If 
$
\cI \cap \mathset{1,2,\dots,n_1}\neq\varnothing
$, 
the fact that
$\sd(H_1)=d(\C_3)$ implies that there is a row of weight one 
in the projection of $H_3$ onto $\cI$. Otherwise
$\cI \subset \mathset{n_1{+}1,n_2{+}2,\dots,n_1+n_2}$,
and the same conclusion follows from\pagebreak[3.99]
$\sd(H_2)=d_2 \ge d(\C_3)$.
\end{proof}

\begin{theorem}
\label{u,u}
Let\/ $\C_1$ be an $(n,k,d)$ binary linear code. Then the code\/
$\C_2 = \{(u,u) : u\in\C_1\}$ 
is a $(2n,k,2d)$ code 
with
\be{thm6}
\rho(\C_2) 
\,\leq\, 
\rho(\C_1)+n
\ee
\end{theorem}
\begin{proof}
Let $H_1$ be a $\rho(\C_1)\!\times n$ parity-check~matrix
for $\C_1$ with $s(H_1) = d$. Construct a parity-check matrix for $\C_2$
as 
$$
H_2
\,=\,
\left(
\begin{array}{@{}c@{\ }c@{}}
H_1 & \zero \\
I_n & I_n
\end{array}
\right)
$$
where $I_n$ is the $n \times n$ identity matrix. Label the
columns of $H_2$ by $0,1,\ldots,2n-1$, and assume to the contrary
there exists a set $\cI \subset \mathset{0,1,\ldots,2n-1}$ 
such that $\abs{\cI} \leq 2d-1$ and there is no row of weight 
one in the projection of $H_2$ onto $\cI$. Let $H_2(\cI)$
denote this projection.
First note that the two identity matrices 
in $H_2$
imply that if $j\in\cI$, then also $(j+n)\!\mod 2n$
is in $\cI$, since otherwise $H_2(\cI)$ contains a row 
of weight one.
It follows that
$\cI\cap\mathset{0,1,\dots,n{-}1}\neq\varnothing$. But 
$\sd(H_1)=d$,~so~the top part of $H_2$ implies that 
$\abs{\,\cI\cap\mathset{0,1,\dots,n{-}1}}\geq d$,
other\-wise $H_2(\cI)$ again contains a row 
of weight one. By the first observation,
we now conclude $\abs{\cI}\geq 2d$, a contradiction.
\vspace{1ex}
\end{proof}

Here is an interesting observation about Theorems \ref{u,v} and \ref{u,u}.
It follows from \eq{thm5} and \eq{thm6} that if the
constituent codes are optimal, in the sense that their stopping 
redundancy is equal~to their redundancy, then the resulting code 
is also optimal. This indicates that the bounds in \eq{thm5} and \eq{thm6}
are tight.

In contrast, the innocuous construction of \emph{extending}
a linear code $\C$ by adding an overall parity-check~\cite[p.\,27]{MWS}
appears~to be much more difficult to handle. The next theorem deals 
only with the special case where $d(\C)=3$.

\begin{theorem}
\label{extension}
Let\/ $\C$ be an $(n,k,3)$ binary linear code. Then the 
extended code\/ $\C'$ is an $(n+1,k,4)$ code with
$$ 
\rho(\C') 
\,\leq\,
2\rho(\C)
\,=\,
2r(\C')-2
$$ 
\end{theorem}
\begin{proof}
Let $H$ be an arbitrary $r(\C)\times n$ parity-check matrix for $\C$. 
We construct a parity-check matrix for $\C'$ as follows
$$ 
H'
\,=\,
\left(
\begin{array}{@{\,}c@{\,\ }c@{\,}}
 H  & \zero \\
\oH & \one
\end{array}
\right)
\vspace{-1ex}
$$ 
where $\oH$ is the bitwise complement of $H$, while
$\zero$ and $\one$ are the all-zero and the all-one column 
vectors, respectively. Label the columns in $H'$ by $1,2,\ldots,n+1$,
and let $\cI$ be a subset~of $\{1,2,\ldots,n+1\}$ with $|\cI| \le 3$.
In fact, it would suffice~to~con-sider the case where 
$\cI \subset \{1,2,\ldots,n\}$ and $|\cI| = 3$;
all other cases easily follow from the fact that $s(H) = 3$
by \Tref{dist3}.

Let $H(\cI)$ and $\oH(\cI)$ denote the projections of $H$ and 
$\oH$,~re-spectively, on the three positions in $\cI$. 
If $H(\cI)$ contains a row of weight one, we are done.
If $H(\cI)$ contains a row of weight two, we are also done ---
then the corresponding row in $\oH(\cI)$ has weight one.
But otherwise, the only rows in $H(\cI)$ are $000$ and $111$,
which means that the three columns in $H(\cI)$ are identical,
a contradiction since $d(\C)=3$.
\vspace{1ex}
\end{proof}

\looseness=-1
The construction in \Tref{extension} 
is not optimal. For example, if $\C'$ is the $(8,4,4)$
extended Hamming code, it produces~a~pa-rity-check matrix for $\C'$ 
with $6$ rows. However, $\C'$ is also the Reed-Muller code $\grm(1,3)$ 
for which we give in the next~section a parity-check matrix $H$
with $s(H) = 4$ and only $5$ rows.

\vspace{3ex}
\section{Reed-Muller Codes}
\vspace{-.25ex}
\label{sec:rm}

\noindent
We now focus on the well-known $(u,u+v)$ construction, 
in particular in connection with the recursive definition
of binary Reed-Muller codes. Our goal is to derive a 
constructive upper bound on the stopping redundancy of
$\grm(r,m)$ --- the binary Reed-Muller code of order $r$
and length $2^m$. 

We begin by recalling several well-known facts.
The reader is referred to~\cite[\hspace{-1pt}Chapter\,13]{MWS}
for a proof of all these facts. For all $r = 0,1,\ldots,m$, 
the dimension of $\grm(r,m)$ is $\smash{k = \sum_{i=0}^r\binom{m}{i}}$
and its minimum distance is \mbox{$d = 2^{m-r}$}.
Let $G(r,m)$ be~a~gene\-rator matrix for $\grm(r,m)$.
Then, using the $(u,u+v)$ construction, $G(r,m)$ can~be
defined 
recursively, as follows:
\be{Grm}
G(r,m)
\ \deff\
\left(
\begin{array}{@{\,}c@{\,}c@{\,}}
G(r,m-1) & G(r,m-1) \\
\zero & G(r{-}1,m{-}1)
\end{array}
\right)
\ee
with the recursion in \eq{Grm} being bootstrapped by 
$G(m,m) \,{=}\, I_{2^m}$ 
and 
$G(0,m) = (11\cdots1)$ 
for all $m$. By convention,
the code $\grm(-1,m)$ is the set $\mathset{\zero}$
for all $m$. Then
\be{dual}
\grm(r,m)^\perp
\,=\,
\grm(m-r-1,m)
\ee
for all $m$ and all $r = -1,0,1,\ldots,m$.
It follows from~\eq{dual} that $G(r,m)$ is a parity-check
matrix for $\grm(m{-}r{-}1,m)$, a code with minimum distance $2^{r+1}$. 
Hence every $2^{r+1}-1$ columns of $G(r,m)$ are linearly
independent.

Our objective in what follows is to construct an
alternative parity-check matrix $H(r,m)$ for
$\grm(m{-}r{-}1,m) = \grm(r,m)^\perp$ such that 
$s\bigl(H(r,m)\bigr) = 2^{r+1}$.
Then the number of rows in $H(r,m)$ gives an upper bound 
on the stopping redundancy of $\grm(m{-}r{-}1,m)$ (and the
number of rows in $H(m{-}r{-}1,m)$~is an upper bound 
on the stopping redundancy of $\grm(r,m)$). Here is 
the recursive construction that we will use.\vspace{.75ex}

\noindent
{\bf Recursive Construction A:}
For all positive integers $m$ and~for all $r = 1,2,\ldots,m-2$,
we define:\vspace{.25ex}
\be{Hrm}
\hspace*{-1.5ex}\Gt(r,m)
=
\left(
\begin{array}{@{}c@{}}
\Strut{6ex}{2.5ex}\gtop\\[0.75ex]
\hline\\[-1.95ex]
\gbot
\end{array}
\right)
\!\,\deff\!\,
\left(
\begin{array}{@{\,}c@{\hspace{-1ex}}c@{}}
\Gt(r,m-1) & \Gt(r,m-1)  \\[.25ex]
\zero & \Gt(r{-}1,m{-}1) \\[0.75ex]
\hline
& \\[-1.75ex]
\Gt(r{-}1,m{-}1) & \zero
\end{array}
\right)
\ee
with the recursion in \eq{Hrm} being bootstrapped as follows:
for all $m = 0,1,\ldots$, the matrices $\Gt(0,m)$, $\Gt(m{-}1,m)$, $\Gt(m,m)$
are defined by\vspace{-1ex}
\begin{eqnarray}
\label{bootstrap1}
\Gt(0,m) 
&\hspace{-1.25ex}\deff\hspace{-1.25ex}& 
G(0,m) = (11\cdots1)\\
\label{bootstrap2}
\Gt(m{-}1,m) 
&\hspace{-1.25ex}\deff\hspace{-1.25ex}& 
G(m{-}1,m)\\
\label{bootstrap3}
\Gt(m,m) 
&\hspace{-1.25ex}\deff\hspace{-1.25ex}& 
G(m,m) = I_{2^m}\\[-2ex]
\nonumber
\end{eqnarray}

\begin{proposition}
\label{generator}
$\Gt(r,m)$ is a generator matrix for $\grm(r,m)$ and,
hence, a parity-check matrix for $\grm(m-r-1,m)$.
\vspace{-.3ex}
\end{proposition}

\begin{proof}
The proof is by induction on $m$ and $r$. 
Equations \eq{bootstrap1} to \eq{bootstrap3} 
establish the induction base. For the induction~step, 
we need to prove that 
\eq{Hrm} generates $\grm(r,m)$, assuming\pagebreak[3.99]
that $\Gt(r,m{-}1)$ 
generates $\grm(r,m{-}1)$ and $\Gt(r{-}1,m{-}1)$ 
genera\-tes $\grm(r{-}1,m{-}1)$. It follows immediately
from \eq{Grm} that $\gtop$ already generates $\grm(r,m)$.
Thus it would suffice to show that all the rows of $\gbot$
belong to $\grm(r,m)$. To this end, we write
\be{RM-recursive}
\grm(r,m)
\, = \,
\left\{ (u,u+v) ~:~ u \in \C_1,~ v \in \C_2 \right\}
\ee
where $\C_1 = \grm(r,m-1)$ and $\C_2 = \grm(r-1,m-1)$. Observe
that each row of $\gbot$ can be written as
$$
(v,\zero)
\, = \,
(v,v) + (\zero,v)
$$
where $v \in \C_2$. The fact that $(\zero,v) \in \grm(r,m)$
follows immediately from \eq{RM-recursive} for $u = \zero$.
The fact that $(v,v) \in \grm(r,m)$ also follows from \eq{RM-recursive}
in conjunction with the well-known fact that 
$\C_2 \subset \C_1$ (take $u := v$ and $v := \zero$).
Hence all the rows~of~$\gbot$ belong to $\grm(r,m)$, 
and the induction step is complete.
\vspace{1ex}
\end{proof}

It remains to show that the stopping distance of $\Gt(r,m)$ is
indeed $2^{r+1}$. We again prove this by induction on $m$ and $r$.
Let us first establish the induction base. Trivially, the stopping 
distance of $\Gt(0,m)$ is $2$, since $\Gt(0,m)=(11\cdots1)$ by 
\eq{bootstrap3}.

\begin{lemma}
\label{baseend}
The stopping distance of $\Gt(m{-}1,m)$ is $2^m$. 
\end{lemma}
\begin{proof}
The proof is by induction on $m$. We start with $m=1$,
in which case we have $s\bigl(\Gt(0,1)\bigr)=2$, as desired. 
For the~in-duction step, observe that
$$
\Gt(m-1,m)
\,=\,
\left(
\begin{array}{@{\,\,}c@{\,}c@{}}
I_{2^{m-1}} & I_{2^{m-1}} \\[.5ex]
\zero & \Gt(m{-}2,m{-}1)
\end{array}
\right)
$$
The situation here is exactly the same as the one we had 
in the proof of \Tref{u,u}, and the result follows in the
same manner. As in \Tref{u,u},
assume to the contrary
there exists a~set $\cI \subset \mathset{0,1,\ldots,2^m-1}$ 
such that $\abs{\cI} \leq 2^m-1$ and there is no row of weight 
one in the projection of $\Gt(m{-}1,m)$~on~$\cI$.
Then $j \in \cI$ implies that $(j+2^{m-1})\!\mod 2^m$ is 
in $\cI$. Hence 
$\cI \cap \mathset{2^{m-1},\dots,2^m-1} \ne \varnothing$.
But the stopping distance of $\Gt(m{-}2,m{-}1)$ is $2^{m-1}$ 
by induction hypothesis, which imp\-lies
that $\mathset{2^{m-1},\dots,2^m-1} \subseteq \cI$.
By the earlier observa\-tion, this means that 
$\cI=\mathset{0,1,\dots,2^m-1}$, a contradiction.
\vspace{1ex}
\end{proof}

\begin{proposition}
\label{RM-sd}
The stopping distance of $\Gt(r,m)$ is $2^{r+1}$
for all positive integers $m$ and for all $r = 0,1,\ldots,m{-}1$,
\end{proposition}

\begin{proof}
The proof is by induction on $m$ and $r$. \Lref{baseend}
in conjunction with the fact that the stopping 
distance of $\Gt(0,m)$ is $2$ establish the induction base.
For the induction step, let $\cI\subseteq\mathset{0,1,\dots,2^m-1}$
be a set of column indices such that $\abs{\cI}\leq 2^{r+1}-1$. 
We distinguish between three easy cases.\vspace{1ex}

\noindent
{\bf Case\,1:}~ 
$\cI \cap \mathset{0,1,\dots,2^{m-1}{-}1} = \varnothing$.\\[.35ex]
Then $\cI\subseteq\mathset{2^{m-1},2^{m-1}{+}\,1,\ldots,2^m{-}\,1}$.
By induction~hypo\-thesis, the stopping distance of
$\Gt(r,m-1)$ is $2^{r+1}$. Hence~the top row in \eq{Hrm}
implies that the projection of $\Gt(r,m)$ onto $\cI$ 
contains a row of weight one.\vspace{1ex}

\noindent
{\bf Case\,2:}~
$1 \leq \abs{\cI\cap\mathset{0,1,\dots,2^{m-1}-1}}\leq 2^r-1$.\\[.35ex]
By induction hypothesis, $\Gt(r{-}1,m{-}1)$ has a stopping dist\-ance of
$2^r$ . Hence the bottom row in \eq{Hrm}
implies that the~proj\-ection of $\Gt(r,m)$ onto $\cI$ 
contains a row of weight one.\vspace{1ex}

\noindent
{\bf Case\,3:}~
$\abs{\cI\cap\mathset{0,1,\dots,2^{m-1}-1}} \geq 2^r$.\\[.35ex]
Then $|\,\cI \cap \mathset{2^{m-1},2^{m-1}{+}\,1,\ldots,2^m{-}\,1}| \le 2^r-1$,
and we are in a case that is symmetric to either Case\,2 or Case\,1.
\vspace{1.25ex}
\end{proof}

The remaining task is to compute the number of rows in~the 
matrix $\Gt(r,m)$. We denote this number as $g(r,m)$.

\begin{lemma}
\label{count}
For all $r = 0,1,\ldots,m-1$, the number of rows in $\Gt(r,m)$ is 
given by\vspace{-1.25ex}
$$
g(r,m)
\,=\,
\sum_{i=0}^r \binom{m{-}r{-}1 + i}{i} 2^i
$$
\end{lemma}

\begin{proof}
Consider the following generating function
$$
f(x,y)
\,=\,
\sum_{m=0}^{\infty}\sum_{r=0}^m g(r,m+1)\,x^r y^m
$$
Note that $\Gt(m,m+1)=G(m,m+1)$ for all $m \geq 0$,~in~view 
of \eq{bootstrap2}. Hence $g(m,m+1)=2^{m+1}-1$.
Using~the~re\-cursion
$
g(r,m+1)
=
g(r,m)+2g(r-1,m)
$,
which follows immediately from \eq{Hrm}, 
along with this initial condition,
we obtain
\be{gen-rec}
f(x,y)
\,=\,
yf(x,y)+2xyf(x,y)+\sum_{i=0}^{\infty}x^i y^i
\vspace{-1ex}
\ee
Upon rearranging, \eq{gen-rec} becomes
\begin{align}
f(x,y)\, & =\ \frac{1}{1-y(1+2x)} \sum_{i=0}^{\infty}x^i y^i 
\nonumber\\
\label{gen2}
& = \parenv{\sum_{i=0}^{\infty}x^i y^i}
\parenv{\sum_{i=0}^{\infty} y^i\sum_{j=0}^i \binom{i}{j}2^j x^j}
\end{align}
The lemma now follows by observing that 
$g(r,m)$ is the~coefficient of $x^r y^{m-1}$ in \eq{gen2}.
\vspace{1ex}
\end{proof}

\hspace*{-2pt}We 
are now in a position to summarize the results of this~sec\-tion
in the following theorem.

\begin{theorem}
\label{RM-final}
For all $m = 1,2,\ldots$ and for all $r = 0,1,\ldots,m$, the 
stopping redundancy of\, $\grm(r,m)$ is upper bounded by
\be{last}
\rho\bigl(\grm(r,m)\bigr)
\,\leq\,
\sum_{i=0}^{m-r-1} \binom{r+i}{i}\,2^i
\ee
\end{theorem}

\begin{proof}
Follows immediately from \eq{dual}, \Pref{generator}, 
\Pref{RM-sd}, and \Lref{count}.\pagebreak[3.99]
\end{proof}

To see how far the bound of \Tref{RM-final} is from the~(conventional) 
redundancy of Reed-Muller codes, we first
need the following technical lemma.

\begin{lemma}
\label{L:lemma15}
For all positive integers $m$ and all\, $0 \le r \le m\,{-}\,1$, 
we have\vspace{-1.5ex}
\be{lemma15}
\sum_{i=0}^r\binom{m{-}r{-}1+i}{i}2^{r-i}
\ = \
\sum_{i=0}^r\binom{m}{i} 
\vspace{.75ex}
\ee
\end{lemma}
\begin{proof}
Denote the sum $\sum_{i=0}^r\binom{m}{i}$ by $S(m,r)$.
Using~the well-known 
$\binom{m}{i}=\binom{m-1}{i-1}+\binom{m-1}{i}$
recursion, we obtain\hspace{10ex}\vspace{-0.50ex}
$$
S(m,r)
\,=\,
\binom{m{-}1}{r} \,+\, 2\!\sum_{i=0}^{r-1}\binom{m{-}1}{i} 
\vspace{-0.50ex}
$$
and recognize the second term above as $2S(m{-}1,r{-}1)$.
The result now follows by induction on $m$ and $r$.
\vspace{.75ex}
\end{proof}

Using \Lref{L:lemma15},
we can establish a relation between~the~re\-dundancy 
of Reed-Muller codes and their stopping redundan\-cy.
For this, it will be more convenient to work with 
the~dual code $\C=\grm(r,m)^\perp$.
Recall that 
$r(\C) = \sum_{i=0}^r\binom{m}{i}$.
Comparing this to the bound on $\rho(\C)$ in 
\Tref{RM-final}, 
we find that
$$
\rho(\C) 
\,\leq\,
\sum_{i=0}^r \binom{m{-}r{-}1+i}{i}2^i 
\,\le\,
2^r \! \sum_{i=0}^r \binom{m}{i}
\,=\,
2^r r(\C)
$$
where the second inequality follows from \eq{lemma15}.
Therefore,~for any fixed order $r$, the stopping redundancy
of $\grm(r,m)^\perp$~is~at most the redundancy of $\grm(r,m)^\perp$
times a constant. Alternatively, if we take 
$\C=\grm(r,m)$, then~\Tref{RM-final} implies that 
$
\rho(\C) \le d(\C) r(\C) / 2
$.
Thus for any fixed $d(\C)$, the increase 
in redundancy is by a constant factor.

\vspace{.50ex} 
\section{Golay Codes}
\vspace{.25ex}
\label{sec:golay}

\noindent
The $(24,12,8)$ binary Golay code $\cG_{24}$ is arguably the 
most~re\-markable binary block code. It is often used as a benchmark
in studies of code structure and decoding algorithms.

There are several ``canonical'' parity-check matrices~for~$\cG_{24}$,
see~\cite{CFV},\,\cite{SPLAG},\,\cite{V-trellis} and other papers. 
Our starting point~is~the~systematic double-circulant matrix $H_{24}$ given 
in~MacWilliams~and Sloane\,\cite[\hspace{-1.75pt}p.65]{MWS} and shown 
in Table\,\ref{tab:bingolay}.
It can be readily verified that $s\bigl(H_{24}\bigr)=4$, which 
means that $H_{24}$ achieves~only~half~of the maximum possible 
stopping distance.
Curiously, the stopping distance of the two ``trellis-oriented''
parity-check matrices for $\cG_{24}$, given in \cite[p.\,2060]{V-trellis}
and \cite[p.1441]{CFV}, is also~$4$.

Computing the general bounds of Theorems \ref{general} and \ref{lowerbound}
for the special case of $\cG_{24}$ 
produces the extremely weak result:
$$
6 \,\leq\, \rho\bigl(\cG_{24}\bigr) \,\leq\, 2509
$$
Having tried several methods to construct a parity-check~mat\-rix 
for $\cG_{24}$ with stopping distance $8$, our best
result was~achi\-eved using a greedy (lexicographic) computer search.
Specifically, with the $4095$ nonzero vectors of $\cG_{24}$ listed
lexicographically, we iteratively construct the parity-check matrix
$H'_{24}$, at each iteration adjoining to $H'_{24}$ the first vector
on the list with the highest score.
Each vector receives $i$ points to its score for each yet uncovered
$i$-set it covers,~where $i \in \{1,2,\ldots,7\}$
(cf.\,\Tref{lowerbound}).
The resulting matrix is given in Table\,\ref{tab:bingolay}.
Since $H'_{24}$ has only $34$ rows and $\smash{s\bigl(H'_{24}\bigr)\!=8}$,  
it~follows that the stopping redundancy of $\cG_{24}$ is 
at most $34$.\pagebreak[3.99]\vspace*{-.5ex}

\begin{table}[ht]
\caption{%
Two parity-check matrices for the $(24,12,8)$ 
Golay code $\cG_{24}$\vspace*{-2.25ex}}
\label{tab:bingolay}
\renewcommand{\arraystretch}{0.72}
$$
H_{24}
\,=
\left(\hspace{-.25ex}
\begin{array}{*{24}{@{\hspace{0.90ex}}c}}
1 &1 &0 &0 &0 &0 &0 &0 &0 &0 &0 &0 &0 &1 &1 &0 &1 &1 &1 &0 &0 &0 &1 &0 \\
1 &0 &1 &0 &0 &0 &0 &0 &0 &0 &0 &0 &0 &0 &1 &1 &0 &1 &1 &1 &0 &0 &0 &1 \\
1 &0 &0 &1 &0 &0 &0 &0 &0 &0 &0 &0 &0 &1 &0 &1 &1 &0 &1 &1 &1 &0 &0 &0 \\
1 &0 &0 &0 &1 &0 &0 &0 &0 &0 &0 &0 &0 &0 &1 &0 &1 &1 &0 &1 &1 &1 &0 &0 \\
1 &0 &0 &0 &0 &1 &0 &0 &0 &0 &0 &0 &0 &0 &0 &1 &0 &1 &1 &0 &1 &1 &1 &0 \\
1 &0 &0 &0 &0 &0 &1 &0 &0 &0 &0 &0 &0 &0 &0 &0 &1 &0 &1 &1 &0 &1 &1 &1 \\
1 &0 &0 &0 &0 &0 &0 &1 &0 &0 &0 &0 &0 &1 &0 &0 &0 &1 &0 &1 &1 &0 &1 &1 \\
1 &0 &0 &0 &0 &0 &0 &0 &1 &0 &0 &0 &0 &1 &1 &0 &0 &0 &1 &0 &1 &1 &0 &1 \\
1 &0 &0 &0 &0 &0 &0 &0 &0 &1 &0 &0 &0 &1 &1 &1 &0 &0 &0 &1 &0 &1 &1 &0 \\
1 &0 &0 &0 &0 &0 &0 &0 &0 &0 &1 &0 &0 &0 &1 &1 &1 &0 &0 &0 &1 &0 &1 &1 \\
1 &0 &0 &0 &0 &0 &0 &0 &0 &0 &0 &1 &0 &1 &0 &1 &1 &1 &0 &0 &0 &1 &0 &1 \\
0 &0 &0 &0 &0 &0 &0 &0 &0 &0 &0 &0 &1 &1 &1 &1 &1 &1 &1 &1 &1 &1 &1 &1 \\
\end{array}
\hspace{-1ex}\right) 
\vspace{2ex}
$$

$$
H'_{24}
\,=
\left(\hspace{-.25ex}
\begin{array}{*{24}{@{\hspace{0.90ex}}c}}
0 &0 &0 &0 &0 &0 &0 &0 &0 &0 &1 &1 &0 &1 &1 &0 &0 &1 &0 &0 &1 &1 &1 &0 \\
0 &0 &0 &0 &0 &0 &0 &0 &0 &0 &1 &1 &1 &0 &0 &1 &1 &0 &1 &1 &0 &0 &0 &1 \\
0 &0 &0 &0 &0 &0 &0 &0 &1 &1 &0 &0 &0 &0 &0 &1 &0 &0 &1 &1 &1 &0 &1 &1 \\
0 &0 &0 &0 &0 &0 &0 &1 &1 &0 &0 &0 &1 &1 &0 &1 &1 &0 &0 &0 &1 &0 &0 &1 \\
0 &0 &0 &0 &0 &0 &1 &0 &0 &1 &1 &1 &1 &1 &1 &0 &0 &0 &0 &1 &0 &0 &0 &0 \\
0 &0 &0 &0 &0 &0 &1 &1 &0 &1 &0 &1 &1 &0 &0 &1 &1 &1 &0 &0 &0 &0 &0 &0 \\
0 &0 &0 &0 &1 &0 &1 &1 &0 &1 &0 &0 &0 &0 &0 &1 &0 &0 &1 &0 &0 &1 &1 &0 \\
0 &0 &0 &0 &1 &1 &1 &0 &1 &1 &0 &1 &1 &0 &0 &0 &0 &0 &0 &0 &0 &1 &0 &0 \\
0 &0 &0 &0 &1 &1 &1 &1 &0 &0 &0 &0 &1 &0 &0 &0 &1 &0 &1 &0 &0 &0 &0 &1 \\
0 &0 &0 &1 &0 &0 &0 &1 &0 &0 &0 &0 &0 &0 &0 &1 &1 &1 &1 &0 &0 &0 &1 &1 \\
0 &0 &0 &1 &1 &0 &0 &0 &0 &0 &0 &0 &0 &1 &1 &1 &0 &1 &1 &0 &0 &1 &0 &0 \\
0 &0 &0 &1 &1 &0 &1 &1 &1 &0 &0 &1 &0 &0 &0 &0 &0 &1 &1 &0 &0 &0 &0 &0 \\
0 &0 &0 &1 &1 &1 &0 &0 &0 &0 &0 &1 &1 &1 &0 &0 &0 &0 &1 &1 &0 &0 &0 &0 \\
0 &0 &1 &0 &0 &0 &1 &1 &0 &1 &0 &0 &0 &0 &0 &0 &1 &0 &0 &0 &1 &0 &1 &1 \\
0 &0 &1 &0 &1 &1 &1 &1 &1 &0 &0 &0 &0 &0 &1 &0 &0 &0 &0 &0 &0 &0 &1 &0 \\
0 &0 &1 &1 &1 &0 &0 &1 &0 &0 &1 &1 &0 &1 &1 &0 &0 &0 &0 &0 &0 &0 &0 &0 \\
0 &1 &0 &0 &0 &0 &0 &0 &0 &0 &1 &0 &1 &0 &1 &0 &1 &0 &0 &1 &0 &1 &1 &0 \\
0 &1 &0 &0 &0 &0 &0 &0 &1 &0 &1 &1 &0 &1 &1 &0 &1 &0 &0 &0 &0 &0 &0 &1 \\
0 &1 &0 &0 &0 &1 &0 &0 &1 &0 &1 &0 &0 &0 &1 &0 &0 &0 &1 &0 &1 &0 &1 &0 \\
0 &1 &0 &0 &0 &1 &1 &1 &1 &0 &0 &1 &0 &0 &0 &0 &1 &0 &0 &0 &1 &0 &0 &0 \\
0 &1 &1 &0 &0 &1 &0 &0 &1 &0 &0 &0 &0 &0 &1 &0 &1 &1 &0 &1 &0 &0 &0 &0 \\
0 &1 &1 &1 &0 &0 &0 &1 &0 &0 &0 &0 &0 &1 &0 &0 &0 &1 &1 &1 &0 &0 &0 &0 \\
1 &0 &0 &0 &0 &0 &1 &0 &0 &0 &1 &1 &1 &0 &0 &1 &0 &0 &0 &0 &0 &1 &1 &0 \\
1 &0 &0 &0 &0 &1 &0 &0 &0 &0 &1 &1 &0 &1 &1 &1 &0 &0 &1 &0 &0 &0 &0 &0 \\
1 &0 &0 &1 &1 &0 &0 &0 &1 &0 &0 &0 &0 &0 &0 &1 &0 &1 &0 &0 &1 &0 &0 &1 \\
1 &0 &1 &0 &0 &0 &0 &0 &0 &1 &1 &0 &1 &0 &0 &0 &0 &0 &0 &1 &0 &0 &1 &1 \\
1 &0 &1 &1 &0 &0 &0 &0 &0 &0 &0 &1 &0 &0 &1 &1 &0 &0 &0 &0 &1 &1 &0 &0 \\
1 &0 &1 &1 &0 &1 &0 &0 &0 &0 &0 &0 &1 &0 &0 &0 &0 &1 &0 &1 &1 &0 &0 &0 \\
1 &0 &1 &1 &1 &0 &0 &0 &0 &0 &0 &0 &0 &1 &0 &0 &0 &0 &0 &1 &0 &1 &0 &1 \\
1 &1 &0 &0 &0 &1 &1 &0 &0 &0 &0 &0 &1 &0 &0 &0 &1 &1 &0 &0 &0 &1 &0 &0 \\
1 &1 &0 &0 &1 &0 &1 &0 &0 &1 &1 &0 &0 &0 &0 &0 &0 &0 &0 &1 &0 &1 &0 &0 \\
1 &1 &0 &1 &0 &0 &0 &0 &0 &1 &0 &0 &0 &1 &0 &0 &0 &1 &0 &0 &1 &1 &0 &0 \\
1 &1 &1 &0 &1 &1 &0 &0 &1 &0 &0 &0 &0 &0 &0 &0 &0 &0 &0 &0 &1 &1 &0 &0 \\
1 &1 &1 &1 &0 &0 &0 &0 &1 &1 &0 &0 &0 &0 &0 &1 &0 &0 &0 &1 &0 &0 &0 &0
\end{array}
\hspace{-1ex}\right)
$$
\end{table}

\begin{table}[ht]
\caption{Number of undecodable erasure patterns\vspace{-.75ex}
\mbox{by weight $w$ in three decoders for $\cG_{24}$}\vspace*{-2.5ex}}
\label{tab:patterns}
\renewcommand{\arraystretch}{1.18}
$$
\begin{array}{ccccc}
\hline
\hline
\\[-1.75ex]
w & \text{Total Patterns} & \NML & \NH & \NHprime\\[1.25ex]
\hline
\\[-2.00ex]
0       & 1               & 0                & 0             & 0             \\
1       & 24              & 0                & 0             & 0             \\
2       & 276             & 0                & 0             & 0             \\
3       & 2024            & 0                & 0             & 0             \\
4       & 10626           & 0                & 110           & 0             \\
5       & 42504           & 0                & 2277          & 0             \\
6       & 134596          & 0                & 19723         & 0             \\
7       & 346104          & 0                & 100397        & 0             \\
8       & 735471          & 759              & 343035        & 3598          \\
9       & 1307504         & 12144            & 844459        & 82138         \\
10      & 1961256         & 91080            & 1568875       & 585157        \\
11      & 2496144         & 425040           & 2274130       & 1717082       \\
12      & 2704156         & 1313116          & 2637506       & 2556402       
\\[0.50ex]
\geq 13 & \binom{24}{w}   & \binom{24}{w}    & \binom{24}{w} & \binom{24}{w} 
\\[1.00ex]
\hline
\hline
\\[-5.00ex]
\end{array}
$$
\end{table}

To evaluate the effect of increasing the stopping distance,
it would be interesting to compare the performance of iterative decoders
for $\cG_{24}$ based on $H_{24}$ or $H'_{24}$, respectively.
As~a~baseline for such a comparison, it would be also useful
to have the performance of a maximum-likelihood decoder for $\cG_{24}$.~In 
what follows, we give analytic expressions for the 
performance of the three decoders on the binary erasure 
channel\,(BEC).

\begin{figure}[t]
\begin{center}
\psfrag{MaximumXLikelihoodXDecoder}%
{\scriptsize \hspace{-1.25ex} maximum-likelihood decoder}
\psfrag{IterativeXDecoderXwithXhf1111}%
{\scriptsize \hspace{-1.25ex} iterative decoder based upon $H'_{24}$}
\psfrag{IterativeXDecoderXwithXhf2222}%
{\scriptsize \hspace{-1.25ex} iterative decoder using $H_{24}$}
\includegraphics[scale=0.71]{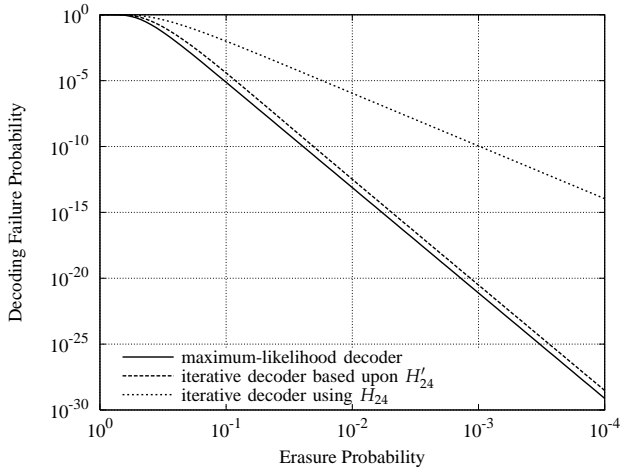}\vspace*{-2ex}
\end{center}
\caption{\hspace*{-1.0ex}The decoding failure probability 
of three decoders for $\cG_{24}$: a~max\-imum-likelihood decoder
and iterative decoders based upon $H_{24}$ and $H'_{24}$}
\label{fig:golay}
\end{figure}

Clearly, a maximum-likelihood decoder fails to decode~(recover)
a given erasure pattern if and only if this pattern~contains the support
of (at least one) nonzero codeword of $\cG_{24}$. Let $\NML$ denote
the number of such erasure patterns as a~function of their weight $w$.
Then 
$$
\Pr{}_{\text{ML}}\{\text{decoding failure}\} 
\, =
\sum_{w=0}^{24} \! \NML\, p^w (1{-}p)^{24-w}
$$
where $p$ is the erasure probability of the BEC.
In contrast, an iterative decoder (based on $H_{24}$ or $H'_{24}$)
fails if and only~if~the 
erasure pattern contains a stopping set. Thus
\begin{eqnarray*}
\Pr{}_{H_{24}}\{\text{decoding failure}\} 
&\hspace*{-1.25ex} = \hspace*{-1.50ex}&
\sum_{w=0}^{24} \! \NH\, p^w (1{-}p)^{24-w}
\\
\Pr{}_{H'_{24}}\{\text{decoding failure}\} 
&\hspace*{-1.25ex} = \hspace*{-1.50ex}&
\sum_{w=0}^{24} \! \NHprime\, p^w (1{-}p)^{24-w}
\end{eqnarray*}
where\hspace{-.75pt} 
$\NH$ and $\NHprime$ denote the number of erasure~patterns
of weight $w$ that contain a stopping set of $H_{24}$~and~$H'_{24}$, 
respectively. It remains to compute $\NHi$, $\NHprimei$, and $\NMLi$.

Obviously, $\NML = 0$ for $w \le 7$ and $\NML = \binom{24}{w}$ for
$w \ge 13$ (any $13$ columns of a parity-check matrix for~$\cG_{24}$
are linearly dependent). For the other values of $w$, we have
$$
\NML
\, = \,
\left\{\hspace*{-.75ex}
\begin{array}{l@{\hspace{4ex}}l}
\displaystyle\binom{16}{w{-}8} 759 & 8 \le w \le 11\\[3ex]
1771(20+720) + 2576   & w = 12
\end{array}
\right.
$$
where we made use of Table\,IV of~\cite{CS} (for $w=12$, 
we have $\NML = |X_{12}| + |S_{12}| + |U_{12}|$ in 
the notation of~\cite{CS}).~To find $\NHi\hspace{-0.25ex}(\cdot)$ 
and $\NHprimei\hspace{-0.23ex}(\cdot)$, we used exhaustive computer 
search. These functions are given in Table\,\ref{tab:patterns}.
The resulting probabili\-ties of decoding failure are plotted in Figure\,1.
Note that while we may add rows to $H'_{24}$ to
eliminate more stopping sets,~this
would have negligible effect since the slope of the performance 
curve is dominated by the smallest $w$ 
for which $\NHprime \neq 0$.

\begin{table}[ht]
\caption{Two parity-check matrices for the $(12,6,6)$ 
Golay code $\cG_{12}$\vspace*{-2.25ex}}
\label{tab:tergolay}
\renewcommand{\arraystretch}{0.60}
$$
H_{12}
\, =
\left(\hspace{-.25ex}
\begin{array}{*{12}{@{\hspace{1.15ex}}c}}
1 &0 &0 &0 &0 &0 &0 &1 &1 &1 &1 &1 \\
0 &1 &0 &0 &0 &0 &1 &0 &1 &- &- &1 \\
0 &0 &1 &0 &0 &0 &1 &1 &0 &1 &- &- \\
0 &0 &0 &1 &0 &0 &1 &- &1 &0 &1 &- \\
0 &0 &0 &0 &1 &0 &1 &- &- &1 &0 &1 \\
0 &0 &0 &0 &0 &1 &1 &1 &- &- &1 &0 \\
\end{array}
\hspace{-1.00ex}\right)\vspace{1.5ex}
$$

$$
H'_{12}
\, =
\left(\hspace{-.25ex}
\begin{array}{*{12}{@{\hspace{.81ex}}c}}
0 &0 &0 &0 &0 &1 &1 &1 &- &- &1 &0 \\
1 &1 &- &- &1 &0 &0 &0 &0 &0 &0 &- \\
0 &0 &0 &- &1 &0 &0 &0 &1 &1 &- &- \\
- &- &1 &0 &0 &1 &1 &1 &0 &0 &0 &0 \\
0 &0 &- &0 &0 &1 &0 &0 &- &1 &- &1 \\
1 &- &0 &1 &1 &0 &1 &- &0 &0 &0 &0 \\
1 &1 &0 &0 &0 &0 &1 &1 &- &0 &0 &- \\
0 &1 &0 &1 &0 &1 &0 &0 &1 &1 &1 &0 \\
1 &0 &- &0 &1 &0 &0 &- &0 &1 &- &0 \\
0 &0 &- &1 &1 &0 &1 &0 &0 &0 &- &1 \\
1 &0 &0 &1 &1 &1 &0 &0 &0 &1 &0 &1 \\
0 &0 &1 &0 &1 &1 &0 &1 &1 &1 &0 &0 \\
- &1 &0 &0 &0 &0 &1 &- &0 &1 &1 &0 \\
0 &- &- &1 &0 &0 &- &1 &0 &0 &0 &- \\
0 &0 &0 &0 &1 &1 &- &0 &1 &0 &1 &1 \\
- &0 &1 &1 &0 &1 &0 &0 &- &- &0 &0 \\
0 &- &1 &0 &- &1 &0 &0 &- &0 &1 &0 \\
- &0 &0 &- &1 &0 &0 &- &0 &0 &1 &1 \\
0 &- &0 &- &1 &0 &- &0 &0 &- &0 &1 \\
1 &0 &- &0 &1 &1 &1 &0 &- &0 &0 &0 \\
1 &0 &1 &1 &1 &0 &0 &0 &1 &0 &1 &0 \\
0 &1 &0 &- &- &1 &0 &0 &0 &0 &- &1 \\
\end{array}
\hspace{-1.00ex}\right)
$$
\end{table}

\vspace*{-4.5ex}

\begin{table}[ht]
\caption{Number of undecodable erasure patterns\vspace{-.75ex}
\mbox{by weight $w$ in three decoders for $\cG_{12}$}\vspace*{-2.5ex}}
\label{tab:terpatterns}
\renewcommand{\arraystretch}{1.10}
$$
\begin{array}{ccccc}
\hline
\hline
\\[-1.75ex]
w & \text{Total Patterns} & \NML & \NHt & \NHprimet\\[1.25ex]
\hline
\\[-2.00ex]
0       & 1               & 0                & 0             & 0             \\
1       & 12              & 0                & 0             & 0             \\
2       & 66              & 0                & 0             & 0             \\
3       & 220             & 0                & 20            & 0             \\
4       & 495             & 0                & 150           & 0             \\
5       & 792             & 0                & 456           & 0             \\
6       & 924             & 132              & 758           & 377         
\\[0.50ex]
\geq 7  & \binom{12}{w}   & \binom{12}{w}    & \binom{12}{w} & \binom{12}{w} 
\\[1.00ex]
\hline
\hline
\end{array}
$$
\end{table}

The $(12,6,6)$ extended ternary Golay code $\cG_{12}$ is another famous 
code. A systematic double-circulant parity-check mat\-rix for~$\cG_{12}$ 
is given in~\cite[p.\,510]{MWS}; this mat\-rix is denoted~$H_{12}$ in
Table\,\ref{tab:tergolay}. It is easy to see that $s\bigl(H_{12}\bigr) = 3$,
which~is~again half of the maximum possible stopping distance.
Using greedy lexicographic search, we have constructed 
a parity-check~mat\-rix~$H'_{12}$ with stopping distance $6$ and only 22 rows. 
This~matrix is also given in Table\,\ref{tab:tergolay}. 
The number of undecodable erasure patterns for a maximum-likelihood 
decoder and for the iterative decoders based on $H_{12}$ and $H'_{12}$
is given in Table\,\ref{tab:terpatterns}.

\vspace{2ex}
\section{MDS Codes}
\label{sec:mds}

\noindent 
The last family of codes we investigate are the maximum~distance
separable (MDS) codes. These codes have intricate~algebraic and
combinatorial structure~\cite[Chapter\,11]{MWS}. In~particular,
if $\C$ is an $(n,k,d)$ linear\footnote{%
Throughout this section, we deal with {linear} MDS 
codes only. Henceforth, whenever we say ``an MDS code'' 
we mean a linear MDS code.}  MDS code, then the dual code $\C^\perp$
is also MDS and its distance is \mbox{$d^\perp = k+1 = n-d+2$}.
Moreover, {every} $d$ positions in $\{1,2,\ldots,n\}$ are the 
support~of a codeword of $\C$\pagebreak[3.99] of weight $d$, while {every} 
$d^\perp$ positions support a~codeword of $\C^\perp$ of weight $d^\perp$.
We will use these and other properties of MDS codes to establish 
sharp upper and lower bounds on their stopping redundancy.

\begin{theorem}
\label{mds}
Let\/ $\C$ be an $(n,k,d)$ MDS code 
with $d \,{\ge}\, 2$.~Then
\be{MDS}
\frac{1}{d-1}\binom{n}{d{-}2} 
\,\leq\
\rho(\C) 
\ \leq\, 
\binom{n}{d{-}2}
\vspace{.75ex}
\ee
\end{theorem}

\begin{proof}
The lower bound 
is just a special case~of~\Tref{lowerbound}.
Taking $i = d-1$ in \eq{w_i}, we find that 
$$
w_{d-1} =\, d(\C^\perp) \,=\, n-d+2
$$
whenever $d \,{\ge}\, 2$, so that $n-w_{d-1} = d-2$.
The \mbox{corresponding} lower bound in \Tref{lowerbound} 
thus reduces to
\be{mds-aux1}
\rho(\C)
\, \geq \,
\frac{\displaystyle \binom{n}{d{-}1}}%
{\displaystyle (n-d+2)\binom{d{-}2}{d{-}2}}
\ = \
\frac{1}{d-1}\binom{n}{d{-}2} 
\ee
To prove the upper bound, note that
every $d^\perp\,{=}\,n-d+2$~po\-sitions support 
a codeword of $\C^\perp$. We take one such codeword
of~$\C^\perp$ for every set of $d^\perp$ 
positions,~and use the resulting 
$$
\binom{n}{d^\perp}
\, = \,
\binom{n}{n\,{-}\,d\,{+}\,2}
\, = \,
\binom{n}{d{-}2}
$$
codewords as rows of a matrix $H$. We claim that $H$ 
is a parity-check matrix for $\C$,
namely that $\rank(H) = n-k = d-1$.
Indeed, consider a set of $d-1$ positions, say
$\{1,2,\ldots,d{-}1\}$. For each $i \in \{1,2,\ldots,d{-}1\}$,
there is a row of $H$ of weight 
$d^\perp = n-(d{-}1)+1$ such that the intersection
of its support with $\{1,2,\ldots,d{-}1\}$ is 
$\{i\}$. The corresponding $d\,{-}\,1$ rows 
of $H$ thus contain an identity matrix on the first
$d\,{-}\,1$ positions; hence $\rank(H) = d-1$.
It remains to show that \mbox{$s(H) = d$}. But 
this follows immediately from what we have
already~prov\-ed: given any set $\cI \subset \{1,2,\ldots,n\}$
with $|\cI| = d\,{-}\,1$, there~is a corresponding set
of $d\,{-}\,1$ rows of $H$ whose projection~on the 
positions in $\cI$ is the identity matrix.
\vspace{1ex}
\end{proof}

Both bounds in \Tref{mds} are exact if $d = 2$. Indeed,~for 
$d=2$ the upper and lower bounds in~\eq{MDS} coincide,
yielding $\rho(\C) = 1$. This reflects the degenerate
case~of~the $(n,n{-}1,2)$ MDS code $\C$, whose dual is
the $(n,1,n)$ repetition code $\C'$. Indeed, any codeword
of $\C'$ can serve as a $1\times n$ parity-check matrix $H$
for $\C$ with $s(H) = 2$. In the case of the
$(n,1,n)$~re\-petition code $\C'$ itself, the bounds in~\eq{MDS}
reduce to
$$
\frac{n}{2}
\ \le\ 
\rho(\C')
\ \le\ 
\frac{n(n{-}1)}{2}
$$
The true value is $\rho(\C') = r(\C') = n-1$. To see this,
consider an $(n{-}1)\times n$ parity-check matrix $H'$
for $\C'$ such that~the~sup\-port of the $i$-th row in $H'$ 
is $\{i,i\,{+}\,1\}$  for $i = 1,2,\ldots,n\,{-}\,1$.\vspace{0.75ex}

Next, we use a combinatorial argument to show that $d=2$
is the \emph{only case} where the lower bound 
of \Tref{MDS} is exact.

\begin{theorem}
\label{mds-lower1}
Let\/ $\C$ be an $(n,k,d)$ MDS code 
with $d \,{\ge}\, 3$. Then
\be{eq:mds-lower}
\rho(\C) 
\ \geq\, 
\floorenv{\frac{1}{d-1}\binom{n}{d{-}2}} \ +\ 1
\vspace{.50ex}
\ee
\end{theorem}

\begin{proof}
Assume to the contrary that there is a parity-check
matrix $H$ for $\C$ with $s(H)\,{=}\,d$ and at most\pagebreak[3.99]
$\binom{n}{d{-}2}/(d\kern0.5pt{-}1)$~rows.
As in \Tref{lowerbound}, we say that a given set 
$\cI \,{\subseteq}\, \{1,2,\ldots,n\}$ with $|\cI| = i$ 
is an \emph{$i$-set}, and that a row $\hhh$ of $H$
\emph{covers} an $i$-set $\cI$ if the projection of $\hhh$
on $\cI$ has weight one. The number~of $(d{-}1)$-sets
covered by a single row of weight $w \ge d^\perp$ is
\be{i-set-cover}
D_{n,d}(w)
\, = \,
\left\{\hspace*{-.75ex}
\begin{array}{l@{\hspace{4ex}}l}
\displaystyle
w \binom{n{-}w}{d-2} & w = d^\perp\kern1pt{=}\, n\,{-}\,d\,{+}\,2 \\[2.0ex]
\hspace*{1.5ex}0     & w > d^\perp\kern1pt{=}\, n\,{-}\,d\,{+}\,2 
\end{array}
\right.
\ee
The total number of $(d{-}1)$-sets is $\binom{n}{d{-}1}$ and
every one of~them must be covered by at least 
one row of $H$. But
\be{Steiner-condition}
\frac{\displaystyle \binom{n}{d{-}1}}%
{\displaystyle \max_{w\kern1pt \ge\kern1.5pt d^\perp} D_{n,d}(w)}
\, = \,
\frac{\displaystyle \binom{n}{d{-}1}}%
{\displaystyle d^\perp\binom{d{-}2}{d{-}2}}
\ = \
\frac{1}{d-1}\binom{n}{d{-}2} 
\ee
in view of~\eq{i-set-cover}. It now follows from \eq{Steiner-condition}
that there are {exactly} $\binom{n}{d{-}2}/(d\kern0.5pt{-}1)$
rows in $H$, all of weight $w = d^\perp$, and that each
$(d{-}1)$-set is covered by \emph{exactly one} row of $H$.
The latter~con\-dition is equivalent to saying that 
each (complementary) set of $n-(d{-}1) = d^\perp\,{-}\,1$
positions is contained in~the~support~of exactly one row 
of $H$. In other words, the supports of the rows of $H$
form an $\cS(d^\perp{-}1,d^\perp,n)$ Steiner system.\footnote{%
An $\cS(t,k,v)$ Steiner system is a set of $k$-subsets 
of $\{1,2,\ldots,v\}$,~called \emph{blocks}, so that each
$t$-subset of $\{1,2,\ldots,v\}$ is contained in exactly 
one~block.}

Such a Steiner system may or may not exist. If it does~not
exist we are done, but in many known cases (e.g.\ 
$\cS(2,3,7)$, $\cS(3,4,8)$, $\cS(4,5,11)$, etc.)
it does; hence we must proceed to establish another
contradiction. To this end, consider a
$(d{-}2)$-set which is the complement of the support 
of a given row $\hhh_1$ of $H$. As $s(H) = d$, this 
$(d{-}2)$-set must be covered~by~some other row of $H$,
say $\hhh_2$. But then
$$
\Bigl|\supp(\hhh_1) \cap \supp(\hhh_2)\Bigr|
\, = \
d^\perp-1
$$
The above means that there is a set of $d^\perp\,{-}\,1$
positions that is contained in two different blocks of 
the $\cS(d^\perp{-}1,d^\perp,n)$ Steiner system, a contradiction.
\vspace{1.25ex}
\end{proof}

{\bf Example.}
The hexacode $\cH_6$ is a remarkable $(6,3,4)$ MDS code
over $\Ffour = \{0,1,\w,\bw\}$. It is unique up to 
monomial equivalence and self-dual under the Hermitian
inner product (so the conjugate of a parity-check 
matrix for $\cH_6$ is a generator matrix for $\cH_6$). 
The upper and lower bounds in~\eq{MDS} 
imply that
$
5 \le \rho(\cH_6) \le 15
$.
Using one of the covering designs (see below) in~\cite{GKP},
we construct the following parity-check~matrix
\renewcommand{\arraystretch}{.90}
\be{hexacode}
H
\, = \,
\left(
\begin{array}{@{}c@{\,}c@{\ \ } c@{\,}c@{\ \ } c@{\,}c@{}}
\\[-2ex]
\bw & \w &   0 &  1 &   0 &  1 \\
\bw & \w &   1 &  0 &   1 &  0 \\
  0 &  1 & \bw & \w &   0 &  1 \\
  1 &  0 & \bw & \w &   1 &  0 \\
  0 &  1 &   0 &  1 & \bw & \w \\
  1 &  0 &   1 &  0 & \bw & \w \\
\end{array}
\right)
\ee
for $\cH_6$. It can be easily verified by hand that 
$s(H) = 4$, and therefore $\rho(\cH_6) \le 6$. Finally,
the lower bound of \Tref{mds-lower1} proves that 
$\rho(\cH_6) = 6$. Thus \eq{eq:mds-lower} 
is exact in this case.\vspace{1.25ex}~\hfill$\Box$

In general, it follows from the proof of \Tref{mds-lower1}
that if $\C$ is an $(n,k,d)$ MDS code and $H$ is a parity-check
matrix for $\C$ with $s(H) = d$, then the supports of rows
of weight $d^\perp$ in $H$ form a $(n,d^\perp,d^\perp{-}1)$
covering design. A \emph{$(v,k,t)$ covering~de\-sign} is 
collection of subsets of size $k$ of $\{1,2,\ldots,v\}$,
called blocks, such that every subset of $\{1,2,\ldots,v\}$
of size $t$ is~con\-tained in at least one block 
(changing ``at least one'' to ``exactly one'' thus makes this 
a Steiner system). The smallest number of
blocks in a $(v,k,t)$ covering design is usually de\-noted by
$C(v,k,t)$ and called the \emph{covering number}
(see~\cite{GKPS},\,\cite{MM} and references 
therein). Thus if $\C$ is an $(n,k,d)$ MDS code, then
\be{cover-general}
\rho(\C)
\,\ge\,
C(n,d^\perp,d^\perp{-}1)
\,=\,
C(n,k\,{+}1,k)
\ee
The best general lower bound on the covering number 
dates back to the work of Sch\"onheim~\cite{Schonheim},
who showed~in~1964~that
$
C(v,k,t) \ge (v/k)\, C(v\,{-}\,1,k\,{-}\,1,t\,{-}\,1)
$.
\hfill For the special case of \eq{cover-general}, this 
proves that
\be{cover1}
\rho(\C)
\,\ge\,
\ceilenv{\frac{n}{k{+}1} 
\ceilenv{\frac{n{-}1}{k} 
\ceilenv{\frac{n{-}2}{k{-}1} 
\cdots
\ceilenv{\frac{n{-}k{+}1}{2}}
\cdots}}}
\ee
Notice that if we ignore all the ceilings in \eq{cover1},
then we recover precisely the lower bound in \eq{MDS}.
Hence \eq{cover1} is always at least as strong as the
lower bound of \Tref{mds}. An alternative 
bound on the covering number is due to de\,Caen~\cite{deCaen}
(see also \cite[p.\,270]{GKP}). In our case, this bound reduces to
\be{cover2}
\rho(\C)
\,\ge\,
\frac{k+1}{(k+2)(d-2)}\binom{n}{d{-}2} 
\ee
This is better than the lower bound of \Tref{mds} if
and only if $n > 2(k\,{+}\,1) = 2d^\perp$.
Note that \Tref{mds-lower1} is sometimes stronger
than both \eq{cover1} and \eq{cover2}, for example
in those cases where 
$n \le 2d^\perp$ and an $\cS(k,k\,{+}\,1,n)$ Steiner 
system exists.\vspace{0.50ex}

We can now summarize most of the results in this section
as follows. If $\C$ is an $(n,k,d)$ MDS code over $\Fq$ 
with $d \ge 3$, the the stopping redundancy of $\C$ is 
in the range
$$
{\frac{1}{d-1}\binom{n}{d{-}2}} 
\ < \, 
\rho(\C) 
\ \leq\, 
\frac{\max\{d^\perp,d{-}1\}}{n}\binom{n}{d{-}2} 
$$
(see Appendix for a proof of the upper bound).
These 
bounds are reasonably close
and, notably, do not depend on the size of the 
field. Determining the stopping redundancy of 
MDS codes \emph{exactly} appears to be a~difficult
combinatorial problem. In view of~\eq{cover-general},
it is likely to be at least as difficult as the
problem of determining~the~cov\-ering number $C(n,k{+}1,k)$.

\vspace{2ex}
\section{Discussion and Open Problems}
\vspace{-.25ex}
\label{sec:conclusion}

\noindent
This paper only scratches the surface of the many interesting and
important problems that arise in the investigation~of~stopping 
redundancy. The importance of stopping sets is well understood
in the case of the binary erasure channel. However, the
concept of stopping redundancy is new. Figure\,1 
clearly~de\-monstrates that it is the stopping sets of size 
strictly less than the minimum distance that are responsible for 
the~per\-formance gap between maximum-likelihood and iterative~decoding. 
Thus eliminating such stopping sets is what we
need~to do, and the stopping redundancy is the relevant figure
of merit. 

It would be extremely interesting to understand how relevant 
stopping redundancy is for other channels. In this regard, 
it is worth mentioning the following observation of 
Feldman~\cite[p.\,176]{Feldman}. In the general framework
of LP decoding, the support of \emph{any pseudocodeword}
is a stopping set for \emph{any channel}. Thus the stopping
redundancy might be the relevant figure of merit in this,
very general, context as well.

\looseness=-1
It is interesting to note that although we have defined and
studied the stopping redundancy as a property of linear codes, it 
turns out to be closely related to a number of well-known
combinatorial structures. Steiner systems 
and covering designs were already discussed in Section\,\ref{sec:mds}. 
A combinatorial structure equivalent to a covering design
is the {Tur\'an system}. For more information on this,
we refer the reader to~\cite{HS},\,\cite{MM},\,\cite{Sidorenko}.
Another combinatorial concept that is \emph{very} closely 
related to stopping redundancy is that of $k$-locally-thin families.
A family $\cF$ of subsets of the set $\{1,2,\ldots,\rho\}$ is said 
to be \emph{$k$-locally-thin} if given any $k$ distinct subsets in $\cF$,
there is at least one element $i \in \{1,2,\ldots,\rho\}$ that is 
contained in exactly~one~of~them. 
The central problem in the study of $k$-locally-thin families
is to determine $M(\rho,k)$, defined as the maximum cardinality 
of a~$k$-locally-thin family of subsets of $\{1,2,\ldots,\rho\}$.
In particular, one would like to determine the sequence
\be{t(k)}
t(k)
\ \ \deff \ \
\limsup_{\rho \to \infty}
\frac{\log_2 M(\rho,k)}{\rho}
\ee
But $M(\rho,k)$ is also the maximum number of columns~in~a~bi-nary matrix 
$H$ with $\rho$ rows, distinct columns,
and no stopping set of size $k$.
Hence, results on stopping redundancy~might 
be relevant in the study
of $k$-locally-thin families, and vice versa.
For example, our construction
in Section\,\ref{sec:rm} produces a parity-check matrix 
for the Reed-Muller code~$\grm(m-2,m)$~of~length $n \,{=}\, 2^m\!$, 
distance $4$, 
and stopping redundancy $\bigl(2+o(1)\bigr)\!\log_2\! n$, thereby
showing that $t(3)\,{\geq}\, 1/2$. It should be pointed out~that
estimating $t(k)$
is a notoriously difficult 
task. In fact,~it~is 
not even known whether $t(3) < 1$ and whether 
$t(k)$~decreases~mo- notonically with $k$. For much more on this,
see \cite{AFK},\,\cite{AKM},\,\cite{Kostochka}, and references therein.

We have concluded the original version of this paper with 
a~variety of research questions related to our results.
Although some of these questions have been since answered
(see below), we repeat them here.
In Section\,\ref{sec:general}, we derived upper and lower
bounds on the stopping redundancy of general binary linear 
codes. Can these general bounds be improved? In particular,
is there an asymptotically good family of codes such that
their stopping redundancy grows only polynomially fast
with their length?
In Section\,\ref{sec:constructions}, we have examined only a small 
sample  of the multitude of known ways of combining~codes~to~con\-struct 
other codes. What can be said of the stopping redundan\-cy
of other constructions, in particular constructions involving 
nonbinary alphabets, such as concatenated/multilevel coding?
In Sections \ref{sec:rm} and \ref{sec:golay}, we investigated 
the Reed-Muller codes and the Golay codes. Are the
constructions provided therein optimal?  In particular,
is it true that $\rho(\cG_{24}) = 34$?
It appears that proving lower bounds
on the stopping redundancy, even for specific codes
such as $\cG_{24}$, is quite difficult.
Finally, in Section\,\ref{sec:mds}, we considered MDS codes.
We conjecture that the stopping redundancy of an $(n,k,d)$
MDS code $\C$ over $\Fq$ \emph{does not depend on the code},
but only on its parameters $n$ and $k$. In other words, any
two $(n,k,d)$ MDS codes have the same stopping redundancy.
If this conjecture is true, then it should be possible, 
in principle, to determine the stopping redundancy 
of an $(n,k,d)$ MDS code as a function of $n$ and $k$. 
However, this appears to be a difficult combinatorial 
problem.\pagebreak[3.99]

\looseness=-1
Finally, we would like to mention two recent papers that
are directly inspired by our results, and improve upon 
them. Etzion~\cite{Etzion} studies in detail the stopping
redundancy of binary Reed-Muller codes. He proves 
that the stopping redundancy of $\grm(m-2,m)$, 
which is also the exteded Hamming code~of length $2^m\!$, 
is $2m-1$. This shows that our construction in Section\,\ref{sec:rm}
is optimal in this case. However, it turns out that this construction 
is \emph{not} optimal for the first-order Reed-Muller codes
$\grm(1,m)$; Etzion~\cite{Etzion} derives
a better upper bound on the stopping~redundancy of these codes.
%
Han and Siegel~\cite{HS} use the ``probabilistic method'' to
establish upper bounds on the stopping redundancy of general
linear codes, which improve significantly upon our result in
\Tref{general}. They also prove upper bounds on the stopping
redundancy of MDS codes in terms of Tur\'an numbers, that 
are 
stronger~than~our~\Cref{mds-final}.

\vspace{3ex}
\appendix[An Improved Upper Bound on the Stopping Redundancy of MDS Bounds]
\vspace{.25ex}

\noindent\looseness=-1
In this appendix, we improve the upper
bound in \Tref{mds} using constant-weight codes. 
An\,\emph{$(n,\kern-1pt 4,\kern-1pt w)$\kern-1pt\ constant-weight~code} 
$\cC$ is 
a set of binary vectors of length $n$ and weight $w$,
such~that any two elements of $\cC$ are at 
distance $\ge 4$ from each other. Let $U(n,w,m)$~de\-note 
the largest possible cardinality of a~union of $m$ 
constant-weight codes, each with parameters $(n,4,w)$.

\begin{theorem}
\label{mds-upper}
Let\/ $\C$ be an $(n,k,d)$ MDS code 
with $d \,{\ge}\, 3$. Set $m = \min\{k,n\,{-}\,k\,{-}\,1\}$.
Then
\be{eq:mds-upper}
\rho(\C) 
\ \leq\, 
\binom{n}{d{-}2} \ - \ U(n,d{-}2,m)
\vspace{.50ex}
\ee
\end{theorem}

\begin{proof}
We start as in the proof of \Tref{mds} by constructing a parity-check
matrix $H$ for $\C$ such that the supports\footnote{%
We shall regard the support of a row of $H$ interchangeably 
as a subset of $\{1,2,\ldots,n\}$ or as the corresponding
binary vector of length $n$.} 
of the rows of $H$ are
all the binary vectors of length $n$ and weight $d^\perp = n-d+2$.
Now let $\cC_1,\cC_2,\ldots,\cC_m$ be any $m$ constant-weight codes
with parameters\hspace{-.75pt} $(n,4,n{-}d{+}2)$\hspace{-.25pt}. 
\hspace{-1.00pt}We 
remove from
$H$ all the rows whose supports belong to 
$
\cC_1 \cup \cC_2 \cup \cdots \cup \cC_m
$.
Let $H'$ denote the resulting matrix. Since obviously
$$
U(n,n\,{-}\,d\,{+}\,2,m)
\,=\,
U(n,d\,{-}\,2,m)
$$
the number of rows remaining in $H'$ is given by the right-hand side 
of~\eq{eq:mds-upper}, provided $\cC_1,\cC_2,\ldots,\cC_m$ are chosen 
so~as~to~max\-imize the cardinality of their union. We claim that $s(H') = d$.
To prove this claim, we distinguish between two cases.
\vspace{.75ex}

\noindent
{\bf Case\,1:}
Consider a $(d{-}1)$-set. As shown in the proof of \Tref{mds},
there are some $d-1$ rows in $H$ such that the projection of
their supports on the $(d{-}1)$-set is the $(d{-}1){\times}(d{-}1)$
iden\-tity matrix. Let $\cD \subset \Fn$ denote this set of 
$d-1$ supports. Any two elements of $\cD$ are at distance
exactly $2$ from each other, since $(d{-}1) + (d^\perp{-}1) = n$.
Hence $|\cD \cap \cC_i| \le 1$ for~all~$i$.~As  
$m \le n\,{-}\,k\,{-}\,1 = d-2 = |\cD| - 1$, it follows that
$H'$ contains at least one row whose support belongs to $\cD$.
\vspace{.75ex}

\noindent
{\bf Case\,2:}
Consider a $t$-set with $t \le d\,{-}\,2$ and assume~w.l.o.g.\
that this $t$-set is $\{1,2,\ldots,t\}$. Note that $H$ 
contains some $d^\perp$ rows whose supports are
$$
\{t,t\,{+}\,1,\ldots,t\,{+}\,d^\perp\}\setminus\{t+i\}
\hspace{4ex}
\text{for $i = 1,2,\ldots,d^\perp$}
$$
As before, let $\cD$ denote this set of $d^\perp$ supports.
The intersection of each support in $\cD$ with the $t$-set
$\{1,2,\ldots,t\}$~is~$\{t\}$, so the projection of each of
the corresponding $d^\perp$ rows of $H$ onto this $t$-set
has weight one. Moreover, any two elements of $\cD$ are,
again, at Hamming distance $2$ from each other. Hence 
$|\cD \cap \cC_i| \le 1$ for all~$i$, and since
$m \le k = d^\perp{-}1 = |\cD| - 1$,~it follows that
$H'$ has at least one row whose support is in $\cD$.
\vspace{.75ex}

It remains to show that $\rank(H') = n-k = d-1$.
But~this follows from the fact that $s(H') = d$.
Indeed, up to an appro\-priate column permutation,
there is a row in $H'$ such that the intersection 
of its support with $\{1,2,\ldots,d{-}1\}$ is $\{d{-}1\}$.
Then, there is another row in $H'$ such that the 
intersection~of~its support with
$\{1,2,\ldots,d{-}2\}$ is $\{d{-}2\}$,
again up to a column permutation. 
Continuing in this manner, we get a set of $d-1$ 
rows of $H'$ whose projection on the first
$d-1$ positions~is~an upper-triangular matrix
with nonzero entries on the main~dia\-gonal.
Hence $\rank(H') = d-1$, and we are done.
\vspace{0.25ex}
\end{proof}

\begin{proposition}
\label{Graham-Sloane}
For all positive integers $n$ and $w$ with $w \le n$ 
and for all $m \le n$,\vspace{-.50ex}
\be{GS-bound}
U(n,w,m) 
\ \ge \
\frac{m}{n} \binom{n}{w}
\vspace{.50ex}
\ee
\end{proposition}

\begin{proof}
Graham and Sloane~\cite[Theorem\,1]{GS} construct~a~partition 
of the set of binary vectors of length $n$ and weight $w$ 
into $n$ constant-weight codes with parameters $(n,4,w)$.
Taking the $m$ largest codes in such a partition proves
\eq{GS-bound}.
\vspace{0.25ex}
\end{proof}

\begin{corollary}
\label{mds-final}
Let\/ $\C$ be an $(n,k,d)$ MDS code. Then
\be{eq:mds-final}
\rho(\C) 
\ \leq\, 
\frac{\max\{d^\perp,d{-}1\}}{n}\binom{n}{d{-}2} 
\vspace{.50ex}
\ee
\end{corollary}

\begin{proof}
Follows immediately from \Tref{mds-upper} and \Pref{Graham-Sloane}.
Note that \eq{eq:mds-final} coincides with \eq{MDS} iff $d=2$.
\vspace{0.75ex}
\end{proof}

\section*{Acknowledgment}
\vspace{-.5ex}

\noindent
We are grateful to Ilya Dumer, Tuvi Etzion, Jonathan Feldman,
and Paul Siegel for helpful and stimulating discussions.




\end{document}